\documentclass[conference]{IEEEtran}
\def\IEEEtitletopspace{18pt}
\IEEEoverridecommandlockouts

\usepackage[utf8]{inputenc}
\usepackage{graphicx}
\graphicspath{ {images/} }

\newcommand{\comment}[1]{}

\newcommand{\be}[0]{\begin{equation}}
\newcommand{\ee}[0]{\end{equation}}
\newcommand{\ben}[0]{\begin{equation*}}
\newcommand{\een}[0]{\end{equation*}}
\newcommand{\bena}[0]{\begin{eqnarray*}}
\newcommand{\eena}[0]{\end{eqnarray*}}
\newcommand{\bea}[0]{\begin{eqnarray}}
\newcommand{\eea}[0]{\end{eqnarray}}

\newcommand{\tends}{\rightarrow}

\IEEEoverridecommandlockouts
\usepackage{cite}
\usepackage{amsmath,amssymb,amsfonts}
\usepackage{algorithmic}
\usepackage{graphicx}
\usepackage{textcomp}
\usepackage{xcolor}
\usepackage[version=4]{mhchem}
\usepackage{siunitx}
\usepackage{array}
\usepackage{longtable}
\usepackage{nomencl}
\usepackage{mathrsfs} 
\usepackage{mathtools} 
\usepackage{optidef}
\usepackage{multicol}
\usepackage{svg}
\usepackage{lipsum}
\usepackage{calrsfs}
\usepackage{scalerel}
\usepackage{stackengine}
\usepackage{booktabs}
\usepackage{caption}
\usepackage{enumerate}

\usepackage[english]{babel}
\usepackage{amsthm}

\DeclareMathAlphabet{\mathcal}{OMS}{cmsy}{m}{n}
\theoremstyle{definition}
\newtheorem{definition}{Definition}
\newtheorem{theorem}{Theorem}

\newtheorem{assumption}{Assumption}

\begin{document}

\title{Safe and Stable Adaptive Control for a Class of Dynamic Systems}

\author{Johannes Autenrieb$^{1}$ and Anuradha Annaswamy$^{2}$
\thanks{$^{1}$ German Aerospace Center (DLR), Institute of Flight Systems, Department of Flight Dynamics and Simulation, 38108, Braunschweig, Germany.
(email: \texttt{johannes.autenrieb@dlr.de})}
\thanks{$^{2}$ Department of Mechanical Engineering, Massachusetts Institute of Technology, Cambridge, MA 02139, USA.
(email: \texttt{aanna@mit.edu}) 
\newline The second author would like to acknowledge the support of the Boeing Strategic University Initiative.}
}

\maketitle

\begin{abstract}
Adaptive control has focused on online control of dynamic systems in the presence of parametric uncertainties, with solutions guaranteeing stability and control performance. Safety, a related property to stability, is becoming increasingly important as the footprint of autonomous systems grows in society. One of the popular ways for ensuring safety is through the notion of a control barrier function (CBF). In this paper, we combine adaptation and CBFs to develop a real-time controller that guarantees stability and remains safe in the presence of parametric uncertainties. The class of dynamic systems that we focus on is linear time-invariant systems whose states are accessible and where the inputs are subject to a magnitude limit. Conditions of stability, state convergence to a desired value, and parameter learning are all elucidated. One of the elements of the proposed adaptive controller that ensures stability and safety is the use of a CBF-based safety filter that suitably generates safe reference commands, employs error-based relaxation (EBR) of Nagumo’s theorem, and leads to guarantees of set invariance. To demonstrate the effectiveness of our approach, we present two numerical examples, an obstacle avoidance case and a missile flight control case.
\end{abstract}


\section{Introduction}
\label{Introduction}
The field of adaptive control has focused on providing real-time inputs for dynamic systems through parameter learning and control design using a stability framework \cite{Narendra2005, Ioannou1996, Sastry_1989, Slotine1991, Krstic1995, Ast13}. A different direction of research has been growing in the area of safety-critical systems \cite{ames2016control, nguyen2016exponential, xiao2019control} motivated by the need to provide verifiable guarantees of safe behavior in systems with mixed autonomy. This paper takes a step in combining adaptive control methods with safety-critical methods for a specific class of dynamic systems. In addition to ensuring safety and stability, the proposed adaptive control design also seeks to accommodate magnitude constraints on the control input.

The tools utilized in establishing stability in adaptive control include Lyapunov stability, analytical continuity, and a reference model that establishes a target for the adaptive control system to track. In the case of safety, control barrier functions (CBFs) and the notion of positive invariance are utilized in order to design the exogenous input into the system. The approach we have used in this paper is a careful combination of both sets of tools; a CBF-based filter design is used in order to design the reference input into the reference model. This in turn is followed by the use of a reference model that utilizes a closed-loop structure \cite{Gibson2013} and a calibration that acknowledges the possibility of input saturation. With the resulting calibrated closed-loop reference model (CCRM), global boundedness and tracking in the absence of input saturation and a domain of attraction result in the presence of input constraints are established. In all cases, safety of the system is guaranteed due to the combined use of Calibrated CBF and CCRM.

Several approaches have been reported in the literature where CBFs are are used to ensure safety for a plant with parametric uncertainties \cite{Xiangru_2015, Jankovic_2018, Buch_2022, Taylor_2020_ml, Ames_2020a, Nguyen_2022}.
Approaches in \cite{Jankovic_2018, Buch_2022} use a robust but conservative approach in the choice of how the system is rendered safe. A less restrictive approach is used in \cite{Taylor_2020_ml} where a CBF filter minimizes the risks of controlling an unknown model and a controller learns how to operate the system via a data-driven approach. A drawback of that approach is that during the learning phase, no guarantees of set invariance or bounds can be provided. In \cite{Ames_2020a}, the notion of a CBF is expanded as an adaptive CBF which requires the barrier function to exist for all unknown parameters and all adaptive gains in a set. A similar approach is proposed in \cite{Nguyen_2022} where L1-adaptive methods are used to update the parameters and the corresponding Control Lyapunov Function (CLF) is assumed to exist for all parameter estimates. While magnitude limits are imposed on the control input in \cite{Nguyen_2022}, those constraints are imposed in the development of the CLFs, which may not be guaranteed to lead to a feasible solution. 

Safety has also been addressed using non-CBF approaches \cite{Lavretsky_2010, Wiese_2017,Muse_2011} in the form of state-constraints. In \cite{Lavretsky_2010,Wiese_2017}, the authors introduced modulation functions that are able to lower the control input such that it never directs the system out of a chosen set of states. In \cite{Muse_2011} a bounding function on the reference input is imposed to lower the control input and therefore ensure state-constraints. The trade offs between the use of such a bounding function and command following are however only empirically addressed in these papers. In contrast, a CBF-based approach, which is used in this paper, allows a streamlined use of a computational solution within a  constrained optimization framework, with a quadratic cost.

In this paper, we propose an adaptive controller that guarantees stability of the closed-loop system in the presence of parametric uncertainties, and safety, both with and without input-saturation. Unlike \cite{Ames_2020a}, the computational burden on CBF is significantly reduced by removing the requirements to be satisfied for all unknown parameters. Instead, the CBF is designed using a reference model and suitably calibrated to accommodate the presence of adaptation.

Preliminaries and problem statement are presented in Section \ref{Preliminaries & Problem Formulation}. The main contributions, the development of an adaptive controller with stability and safety properties, are presented in Sections \ref{Safe_OMRAC} and \ref{Safe_CCRM}. Section \ref{Safe_OMRAC} does not include any magnitude limits on the control input, while Section \ref{Safe_CCRM} includes these limits. Section \ref{Simulation} includes numerical simulations.

\section{Preliminaries \& Problem Formulation}
\label{Preliminaries & Problem Formulation}
\subsection{Preliminaries}
\label{Preliminaries}
We define a nonlinear continuous system
\begin{equation}
    \label{NonlinearSystem}
    \dot{x}(t) = f(x(t))
\end{equation}
where  $ x(t) \in \mathbf{R}^n$. In order to define safety, we consider a continuously differentiable function  $h: \chi \rightarrow \mathbf{R}$ where $\chi \subset \mathbf{R}^n$, and a set $S$ defined as the zero-superlevel set of $h$, yielding:
\begin{equation}
    \label{Safe_set_1}
    S \triangleq \begin{Bmatrix} x(t) \in  \chi | h(x(t)) = 0 \end{Bmatrix}
\end{equation}
\begin{equation}
    \label{Safe_set_1}
    \partial S \triangleq \begin{Bmatrix} x(t) \in  \chi | h(x(t)) \geq 0 \end{Bmatrix}
\end{equation}
\begin{equation}
    \label{Safe_set_3}
    int(S) \triangleq \begin{Bmatrix} x(t) \in  \chi | h(x(t)) > 0 \end{Bmatrix}
\end{equation}

The following definitions are introduced \cite{Nagumo_1942, Blanchini_1999}:
\begin{definition}
The set $S$ is positively  invariant for the system \eqref{NonlinearSystem}, if for every $x_0 \in S$, it follows $x(t) \in S$ for $x(0) = x_0$ and all $t \in I(x_0) = [0,\tau_{max} = \infty)$.
\end{definition}
\begin{definition}
The set $S$ is weakly positively invariant for the system \eqref{NonlinearSystem}, if among all the solutions of \eqref{NonlinearSystem} originating in $x_0 \in S$, there exists at least one globally defined solution $x(t)$ which remains inside $S$ for all $t \in I(x_0) = [0,\tau_{max} = \infty)$.
\end{definition}
Next, we define the distance from a point to a set:
\begin{definition}
\label{distance_definition}
Given a set $S \subset \mathbf{R}^n$ and a point $y \subset \mathbf{R}^n$, the distance from the point to the set is defined as
\begin{equation}
    dist(y,S) = \inf_{w \in S} \| y-w \|_*
\end{equation}
where $\| \cdot \|_*$ is any relevant norm.
\end{definition}
Based on Definition \ref{distance_definition}, we can formulate the definition of a tangent cone for a closed set.
\begin{definition}
\label{tangent_cone_definition}
Given a closed set $S$, the tangent cone to $S$ at $x$ is defined as:
\begin{equation}
    T_S (x) = 
    \begin{Bmatrix} 
    z: \liminf\limits_{\tau \tends 0} \frac{dist(x+\tau z,S)}{\tau}=0
    \end{Bmatrix}
\end{equation}
\end{definition}
If $S$ is convex $T_S(x)$ is convex, and “$\liminf$” can be replaced by “$\lim$”. Furthermore if $x \in int(S)$, then $T_S (x) = \mathbf{R}^n$, whereas if $x \notin S$, then $T_S (x) = \emptyset$, since $S$ is defined as a closed set. Therefore $T_S (x)$ is only non-trivial on the boundary of $S$.

We use Definition \ref{tangent_cone_definition} to introduce Nagumo's theorem\cite{Nagumo_1942}:
\begin{theorem}
\label{Nagumos_theorem_number}
Consider the system defined in \eqref{NonlinearSystem}. Let $S \subset \mathbf{R}^n$ be a closed set. Then, $S$ is weakly positively invariant for the system if and only if \eqref{NonlinearSystem} satisfies the following condition:
\begin{equation}
    \label{Nagumos_theorem}
    f(x(t)) \in T_S (x(t)), \,\,\, \text{for} \,\,\, \forall x \in S
\end{equation}
\end{theorem}
The theorem states that if the direction of the dynamics defined in \eqref{NonlinearSystem} for any $x(t)$ at the boundary of the safe set $\partial S$ points tangentially or inside to the safe set $S$, then the trajectory $x(t)$ stays in $S$. 
\begin{definition}
A continuous function $\alpha: (-b, a) \rightarrow \mathbf{R}$, with $a,b > 0$, is an extended class $\mathcal{K}$ function $(\alpha \in \mathcal{K})$, if $\alpha(0) = 0$ and $\alpha$ is strictly monotonically increasing. If $a, b = \infty$, $lim_{r \rightarrow \infty} \alpha(r) = \infty$, $lim_{r \rightarrow -\infty} \alpha(r) = -\infty$ then $\alpha$ is said to be a class $\mathcal{K}_{\infty}$ function $(\alpha \in \mathcal{K}_{\infty})$.
\end{definition}
\begin{definition}
For the system considered in \eqref{NonlinearSystem}, a continuously differentiable and convex function $h: \mathbf{R}^n \rightarrow \mathbf{R}$ is a zeroing barrier
function (ZBF) for the set $S$ defined by \eqref{Safe_set_1} and \eqref{Safe_set_3}, if there exist an extended class $\mathcal{K}$ function $\alpha(h(x(t)))$ and a set $S \in \mathbf{R}^n$ such that $\forall x \in S$,
\begin{equation}
    \label{ZeroingBarrierFunction}
    \dot{h}(x(t)) \geq -\alpha(h(x(t)))
\end{equation}
\end{definition}
The above definitions lead to a less restrictive version of \eqref{Nagumos_theorem} as it weakens the requirement to that in \eqref{ZeroingBarrierFunction}. 

We now expand the scope of the problem statement from \eqref{NonlinearSystem} to those with an affine control input, of the form:

\begin{equation}
    \label{NonlinearPlant}
    \dot{x} = f(x(t)) + g(x(t)) u(t)
\end{equation}
where $g$ is Lipschitz and  $u(t) \in \mathbf{R}^m$. We introduce the notion of a Control Barrier Function (CBF) such that its existence allows the system to be rendered safe w.r.t. $S$ \cite{Ames_2014,Ames_2017} and allows a weaker requirement for system safety with a control input $u(t)$, similar to \eqref{ZeroingBarrierFunction}.
\begin{definition}
Let $S \subset \chi$ be the zero-superlevel set of a continuously differentiable function $h: \chi \rightarrow \mathbf{R}$. The function $h$ is a zeroing control barrier
function (ZCBF) for $S$, if there exists a class $\mathcal{K}_{\infty}$ function $\alpha(h(x(t)))$ such that for the system defined in \eqref{NonlinearPlant} we obtain:
\begin{equation}
    \label{ControlBarrierFunction}
    \sup_{u\in \mathbf{R}^m} \frac{\partial h}{\partial x}\begin{bmatrix} f(x(t)) + g(x)u(t) \end{bmatrix}  \geq -\alpha(h(x(t)))
\end{equation}
for all $x \in S$.
\end{definition}
Using the Lie derivative notation, we obtain the following formulation for a ZCBF considering the system defined in \eqref{NonlinearPlant}:
\begin{equation}
\dot{h}(x(t)) = L_{f} h(x(t)) + L_{g} h(x(t)) u(t)
\vspace{0.2cm}
\end{equation}

\subsection{Problem Formulation}
\label{Problem Formulation}
We consider a linear plant with parametric uncertainties of the form:
\begin{equation}
\label{LinearPlantModel}
\dot{x}_p(t) = A_p x_p(t) + B_p \Lambda (R_{u_0}(u(t))
\end{equation}
where $x_p(t) \in \mathbf{R}^{n}$ is a measurable state vector and $u(t) \in \mathbf{R}^{m}$ is a control input vector. The matrices $A_p \in \mathbf{R}^{n \times n}$  and $\Lambda \in \mathbf{R}^{m \times m}$ are unkown and $\Lambda$ has only diagonal positive entries. The input matrix $B_p \in \mathbf{R}^{n \times m}$ is known. The control input is assumed to be magnitude limited by $u_0$, which is represented using the function $R_{u_0}(\cdot)$ as
\begin{equation}
\label{magnitude_limits}
    R_{u_0}(u(t)) = \begin{dcases*}
        $$u(t)$$           & if  $\| u(t)\| \leq u_0$ \\
        $$u_0 \frac{u(t)}{\| u(t) \|}$$  & if  $\| u(t) \| > u_0$ 
    \end{dcases*}
\end{equation}
The objectives are to determine a $u(t)$ for \eqref{LinearPlantModel} such that the plant state $x_p(t)$ tracks a desired reference $ x_d(t)$ and that for any initial condition $x_0 := x(t_0) \in S$, it is ensured that the plant state vector $x_p(t)$ stays within the safe set $S \in \mathbf{R}^n$. This is equivalent to having the control input ensures that there is a CBF with $h(x(t)) \geq 0$ for $\forall t \geq 0$.


\section{Safe Adaptive Control Design with Open-Loop Reference Model}
\label{Safe_OMRAC}
In this section, we consider a simpler version of the problem statement, where the control input magnitude limit is removed. The main challenge in ensuring positive invariance for the plant in \eqref{LinearPlantModel}, which has uncertainties in $A_p$ and $\Lambda$, is the design of a suitable ZCBF, which requires the model to be known, as is evident from \eqref{ControlBarrierFunction}. We therefore first choose a target system, i.e. a reference model, that the adaptive system can be made to contract towards. This reference model is chosen so that its state approaches the desired reference $x_d(t)$ and simultaneously allows the generation of a suitable ZCBF that ensures safety. We choose an open-loop reference model (ORM) of the form:
\begin{equation}
    \label{ORM}
    \dot{x}_m(t) = A_m x_m(t) + B_m r(t)
\end{equation}
where $x_m(t) \in \mathbf{R}^{n}$ is the reference model state vector and $r(t) \in \mathcal{R} \subset \mathbf{R}^{m}$ the reference input vector. The matrix $A_m \in \mathbf{R}^{n \times n}$ is a Hurwitz matrix, and $B_m \in \mathbf{R}^{n \times m}$ has full column rank. It is easy to see that if
\begin{equation}
    \label{Reference_signal}
    r(t) = B_m^+ (\dot{x}_d(t) - A_m x_d(t))
\end{equation}
then $x_m(t)$ approaches $x_d(t)$, where $B_m^+$ denotes the Moore-Penrose inverse of $B_m$. 

In order to ensure that $x_m(t)$ stays inside a safe set $S$, rather than choose $r$ as in \eqref{Reference_signal}, we use a QP-ZCBF safety filter for the plant in \eqref{LinearPlantModel} as follows \cite{Xiangru_2015}:
\begin{align}
    \label{LCBF_QP_R}
   &\min_{r(t)\in \mathcal{R}}
   \begin{aligned}[t]
      &\|r - r^*\|_2
   \end{aligned} \\
   &\text{s.t.} \notag \\
   & \frac{\partial h}{\partial x} \begin{bmatrix}
A_m x_m + B_m r \end{bmatrix} \geq - \alpha(h(x_m))+\Delta, \notag
\end{align}
where $r^*= B_m^+ (\dot{x}_d - A_m x_d)$  and $\Delta>0$ is a positive constant that introduces a safety buffer. The optimal solution of \eqref{LCBF_QP_R} can be easily determined by using KKT conditions given by:
\begin{subequations}
\label{KKT_r}
\begin{align}
& r - r^* - L_{B_m} h(x_m)^T \lambda  = 0 \label{KKT_r_1}\\
& \lambda (-L_{A_m x_m} h(x_m) - L_{B_m} h(x_m) r  - \alpha(h(x_m)) + \Delta) = 0\label{KKT_r_2}\\
& -L_{A_m x_m} h(x_m) - L_{B_m} h(x_m) r - \alpha(h(x_m))  + \Delta \leq 0 \label{KKT_r_3}
\end{align}
\end{subequations}
with a suitable choice of $\lambda\geq 0$.

In what follows, for ease of exposition, we choose the class $\mathcal{K}$-function $\alpha(h(x))=\gamma h(x)$, where $\gamma$ is a positive scalar constant \cite{Ames_2014, Zeng_2021}. 

%
%

\subsection{Adaptive control design}
Our reference system is now determined using
\eqref{ORM} and  \eqref{LCBF_QP_R} as 
\begin{equation}
    \label{ORM2}
    \dot{x}_m(t) = A_m x_m(t) + B_m r_s(t)
\end{equation}
The following assumptions are made regarding the unknown parameters in \eqref{LinearPlantModel}:
\begin{assumption}
\label{assumption_matching_condition}
Constant matrices $\theta_x^*$ and $\theta_r^*$ exist that solve the following:
\begin{equation}
\label{matching_condition_1}
A_m = A_p + B_p \Lambda \theta_x^*
\end{equation}
\begin{equation}
\label{matching_condition_2}
B_m = B_p \Lambda \theta_r^*
\end{equation}
\end{assumption}
\begin{assumption}
The uncertainty $\Lambda$ is a diagonal positive definite matrix.
\end{assumption}

We now propose the adaptive controller for the plant in \eqref{LinearPlantModel}:
\begin{equation}
    \label{adaptive_feeback_controller}
    u(t)  = \widehat{\theta}_x(t) x_p(t) + \widehat{\theta}_r(t) r_s(t)
\end{equation}
The time-varying parameters in \eqref{adaptive_feeback_controller} are adjusted using the following adaptive laws: 
\begin{equation}
\label{theta_x_adaption_law}
    \dot{\widehat{\theta}}_x(t) = - \Gamma_x x_p(t) e_x(t)^T P B_p\,, \,\,\,\,\,\, \Gamma_x>0
\end{equation}
\begin{equation}
\label{theta_r_adaption_law}
    \dot{\widehat{\theta}}_r(t) = - \Gamma_r r(t) e_x(t)^T P B_p\,, \,\,\,\,\,\, \Gamma_r>0
\end{equation}
where $e_x(t)=x_p(t)-x_m(t)$ and $P$ is the solution of the Lyapunov equation $A_m^T P + P A_m = -Q$, where $Q>0$. Both $\Gamma_x$ and $\Gamma_r$ are positive definite matrices defined as the adaptive update gains. We further introduce a corresponding output error $e_u(t)=u(t)-u^*(t)$, which will be useful to quantify the safety of the adaptive controller, where $u^*(t)$ represents the ideal control input and is defined as:
\begin{equation}
\label{eq:idealu}
u^*(t)  = \theta_x^* x_p(t) + \theta_r^* r_s(t)
\end{equation}
In what follows, it will be assumed that the reference input $r_s(t)$ has a bounded derivative.
\begin{theorem}
\label{Theorem_Basic_Adaptive_Controller}
The overall closed-loop adaptive system defined by the plant in \eqref{LinearPlantModel}, the control input in \eqref{adaptive_feeback_controller} and the adaptation laws  in \eqref{theta_x_adaption_law}, \eqref{theta_r_adaption_law} has globally bounded solutions for any initial conditions $x_p(t_0)$, $\widehat\theta_x(t_0)$, and $\widehat\theta_r(t_0)$ and both the errors $e_x(t)$ and $e_u(t)$ converge to zero as $t\rightarrow\infty$.
\end{theorem}
The proof of the theorem follows from standard adaptive control arguments, since the error dynamics is of the form
\begin{equation}
    \label{error_dyanmics_1}
    \dot{e}_x(t) = A_m e_x(t) + B_p \Lambda (\widetilde{\theta}_x(t) x_p(t) + \widetilde{\theta}_r(t) r_s(t))
\end{equation} where 
    $\widetilde{\theta}_x(t) = \widehat{\theta}_x(t) - \theta^*_x$, $
    \widetilde{\theta}_r(t) = \widehat{\theta}_r(t) - \theta^*_r$
and together they admit a Lyapunov function 
\begin{equation}
\begin{split}    
    \label{Lyapunov_function1}
    &V(e_x(t), \widetilde{\theta}_x(t), \widetilde{\theta}_r(t)) = \frac{1}{2} e_x^T(t) P e_x(t)\\
   &+ \frac{1}{2} \operatorname{Tr} [ \widetilde{\theta}_x(t) \Gamma_x^{-1} \widetilde{\theta}_x^T(t) \Lambda ] + \frac{1}{2} \operatorname{Tr}  [ \widetilde{\theta}_r(t) \Gamma_r^{-1} \widetilde{\theta}_r^T(t) \Lambda ]
\end{split}
\end{equation}
It is easy to see that  $\dot V \leq 0$, and $e_x \in \mathcal{L}_2$. As $e_x(t)$ is bounded and has a bounded derivative, an application of the Barbalat's Lemma leads to $\lim_{t \to \infty} e_x (t) = 0$ \cite{Narendra2005}. From \eqref{error_dyanmics_1}, it follows that $e_u(t)$ is an input into an LTI system, with a bounded derivative, whose state is $e_x(t)$; therefore it follows that $\lim_{t \to \infty} e_u (t) = 0$ \cite{Annaswamy2021}.

%
%

\subsection{Safety in the presence of uncertainties}

With stability guaranteed from the discussions above, we now derive conditions for the safety of the proposed adaptive controller. The core ideais to render the known reference model to be safe i.e. $h(x_m) \geq 0$ for $\forall t \geq 0$, and to use the adaptive system to make the closed-loop contract towards the reference model, which ensures that $h(x_p) \geq 0$ for $\forall t \geq 0$. The goal is to derive conditions under which 
\begin{equation}
\label{eq:safety}
\dot h_p\geq -\gamma h(x_p(t)) +\Delta
\end{equation}
We note that the QP-CBF filter ensures that
\begin{equation}
    \label{CBF_reference_model}
    \dot{h}_m   = \underbrace{\frac{\partial h}{\partial x}|_{x_m}}_{a_0} 
    [ \underbrace{A_m x_m(t) + B_m r_s(t)}_{a_1}
    ]\geq -\underbrace{\gamma h(x_m(t))}_{a_2}+\Delta
\end{equation}
where $\dot h_p=\frac{\partial h}{\partial x}|_{x_p}$ and $\dot h_m=\frac{\partial h}{\partial x}|_{x_m}$. To ensure that a ZCBF exists for the adaptive system specified by \eqref{LinearPlantModel},\eqref{adaptive_feeback_controller}-\eqref{theta_r_adaption_law}, we consider
\begin{equation}
    \label{CBF_plant_adaptive}
    \begin{split}
    \dot{h}_p = \frac{\partial h}{\partial x}|_{x_p}
    [
    A_p x_p(t) + B_p \Lambda ( \widehat{\theta}_x(t) x_p(t)    + \widehat{\theta}_r(t) r_s(t) ) ]
    \end{split}
\end{equation}
From \eqref{LinearPlantModel}, \eqref{matching_condition_1}-\eqref{adaptive_feeback_controller}, \eqref{eq:idealu}, \eqref{CBF_reference_model}, and the definition of the errors $e_x$ and $e_u$, we obtain that
\begin{equation}
    \label{hp_bo_a1_ep}
    \begin{split}
    \dot{h}_p = 
    \underbrace{\frac{\partial h}{\partial x}|_{x_p}}_{b_0}
    [&\underbrace{A_m x_m(t) + B_m r_s(t)}_{a_1} \\ &+ \underbrace{A_m e_x(t) + B_p \Lambda e_u(t)}_{\Bar{e}}
    ]
    \end{split}
\end{equation}
Algebraic manipulations allow us to rewrite \eqref{hp_bo_a1_ep} using \eqref{CBF_reference_model} as
\begin{equation}
    \label{CBF_deriation_inequality}
    \begin{split}
    \dot{h}_p &= b_0 [a_1 + \Bar{e}]\\ 
    &\geq - a_2 +\Delta+  \underbrace{a_0 \Bar{e} + (b_0 - a_0)\Bar{e} + (b_0 - a_0) a_1}_{z(t)}
    \end{split}
\end{equation}
Since the goal is to establish safety of the closed-loop adaptive system, we utilize the following two inequalities:
\begin{equation}
\label{lipschitz_gradient}
    \begin{split}
    | g(x_p(t)) - g(x_m(t)) | \leq \kappa_1 | e_x(t) |
    \end{split}
\end{equation}
\begin{equation}
\label{lipschitz_barrierfunction}
    \begin{split}
    | h(x_p(t)) - h(x_m(t)) | \leq \kappa_2 | e_x(t) |
    \end{split}
\end{equation}
where $g(x(t))=\frac{\partial h}{\partial x}$, $\kappa_1$ and $\kappa_2$ are Lipschitz constants associated with $g(x_p(t))$ and $h(x_p(t))$, respectively. We note additionally from Theorem \ref{Theorem_Basic_Adaptive_Controller} that $\|x_p(t)\|$, $\|h(x_p(t))\|$ and $\|\Bar{e}\|$ are bounded. Therefore, $|z(t)|\leq z_0$, where $z_0$ is defined as
\begin{equation}
    z_0= |a_0| |\Bar e| + \kappa_1 |e_x(t) | \begin{pmatrix} |\Bar e| + |a_1| \end{pmatrix}
\label{eq:bound1}
\end{equation} 
Using the lower bound $-z_0$ for $z(t)$, we rewrite \eqref{CBF_deriation_inequality} as
\begin{equation}
    \label{lower_bound_h_dot}
    \begin{split}
    \dot{h}_p \geq - \gamma h(x_p(t)) +\Delta- \bar{F}(|\Bar{e}|, | e_x(t) |)
    \end{split}
\end{equation}
where
\begin{equation}
   \bar{F}(|\Bar{e}|, | e_x(t) |) =  \gamma\kappa_2 | e_x(t) | + |a_0| |\Bar e| + \kappa_1 |e_x(t) | \begin{pmatrix} |\Bar e| + |a_1|\end{pmatrix}
\end{equation}
The inequality in \eqref{lower_bound_h_dot} implies that safety of the closed-loop adaptive system will be guaranteed after $t\geq t_0+T$, where $T$ is a finite interval, as $\Bar{F}(t)\rightarrow 0$ as $t\tends \infty$, and therefore $|\Bar{F}(t)|\leq \Delta\; \forall t \geq t_0+T$. This in turn implies that the closed-loop adaptive system will remain safe for all $t\geq t_0$ if
\begin{equation}
h(x_p(t))\geq h_0 \,\,\,\,\, \forall t\geq [t_0,t_0+T]
\label{eq:safetycondns}
\end{equation}
where $\gamma h_0 \geq F_{{\rm max}}$ where 
\begin{equation}
\label{eq:safetycondn_Fs}
F_{{\rm max}} = \max_{t\in [t_0,t_0+T]}\bar{F}(t)
\end{equation}
As $F(t)$ is bounded, it is clear that such an $F_{{\rm max}}$ exists. Condition \eqref{eq:safetycondns} is satisfied if there is a separation between the period of adaptation and the time at which the system approaches its limit of safety. This property is summarized in the following theorem, where $e_h(t):=h(x_p(t))-h(x_m(t))$.
\begin{theorem}
\label{Theorem_Safe_Basic_Adaptive_Controller}
A ZCBF $h(x)$ exists for all $S$ in $\mathbf{R}^n$ for the overall closed-loop adaptive system defined by the plant in \eqref{LinearPlantModel} and the adaptation laws in \eqref{theta_x_adaption_law} and \eqref{theta_r_adaption_law}, if \eqref{eq:safetycondns} is satisfied, with $\gamma h _0\geq F_{{\rm max}}$ where $F_{{\rm max}}$ is defined as in \eqref{eq:safetycondn_Fs}. Further, the inequality \eqref{eq:safetycondns} also implies that $\lim_{t\tends\infty}e_h(t)=0$.
\end{theorem}
The following choice of $\gamma$ as a function of the safety error may allow the condition \eqref{eq:safetycondns} to be satisfied for a larger class of ZCBFs:
 \begin{equation}
    \label{error_based_gamma_1}
    \gamma(e_h(t)) = \gamma_0 e^{-(\epsilon e_h(t))^2}
\end{equation}
with $\gamma_0 \geq 0$ and $\epsilon \geq 0$ are positive constants. Such a choice allows $\gamma$ to take  on a value that is close to $\gamma_0$ as long $e_h$ is small, and $\gamma(e_h(t))$ allows it to become small as $|e_h(t)|$ increases. The rationale for such a choice is that near $t_0$, when the transients of the adaptive system are yet to settle down, $e_h$ may be large and therefore a conservative choice of $\gamma$ near zero is prudent; as time proceeds, the adaptive system ensures that $x_p$ approaches $x_m$, and therefore $e_h$ approaches zero. As this occurs, $\gamma$ can be relaxed to take on larger values. We denote such a choice of Eq. \eqref{error_based_gamma_1} as an error-based relaxation (EBR). It should be noted that the proposed adaptive controller ensures learning in the form of minimization of all performance errors $e_x(t)$, $e_u(t)$, and $e_h(t)$ to zero.
\section{Safe Adaptive Control Design with  Calibrated Closed-Loop Reference Model}
\label{Safe_CCRM}
We now consider the adaptive control of \eqref{LinearPlantModel} subject to the magnitude limit as in \eqref{magnitude_limits}. In order to accommodate these limits and to improve on the transient performance we propose a calibrated closed-loop reference model (CCRM) of the form 
\begin{equation}
    \label{CCRM}
    \dot{x}_m = A_m x_m(t) + B_m r_s(t) + Le_x(t) + B_p \hat{\Lambda}\Delta u(t)
\end{equation}
where $L$ is a matrix such that $(A_m-L)$ is Hurwitz, $\Delta u(t)=R_{u_0}(u(t))-u(t)$ and represents a disturbance due to saturation, and $\hat{\Lambda}$ is an estiamtion of the unknown matrix $\Lambda$. The input $r_s(t)$ is the solution of a modified QP-ZCBF filter, which is defined by the following constrained optimization:
\bea 
\min_{r\in \mathcal{R}} 
      (r - r^*)^2\,\,\,\,\,\,\,\,\,\,\,\,\,\,\,\,\,\,\,\,\,\,\,\,\,\,\,\,\,\,\,\,\,\,\,\,\,\,\,\,\,\,\,\,\,\,\,\,\,\,\,\,\,\,\,\,\,\,\,\,\,\,\,\,\,\,\,\,\,\,\,\,\,\,\,\,\,\,\,\,\,\,\,\,\,\,\,\,\,\,\,\,\,\,\,\,\,\,\,\,&&\\
   \text{s.t.} \hfill \,\,\,\,\,\,\,\,\,\,\,\,\,\,\,\,\,\,\,\,\,\,\,\,\,\,\,\,\,\,\,\,\,\,\,\,\,\,\,\,\,\,\,\,\,\,\,\,\,\,\,\,\,\,\,\,\,\,\,\,\,\,\,\,\,\,\,\,\,\,\,\,\,\,\,\,\,\,\,\,\,\,\,\,\,\,\,\,\,\,\,
   \,\,\,\,\,\,\,\,\,\,\,\,\,\,\,\,\,\,\,\,\,\,\,\,\,\,\,\,\,\,\,\,\,\,\,\,\mbox{}&&\nonumber\\
    \frac{\partial h}{\partial x} \left[
A_m x_m + B_m r \right.\left.+ L R_{e_0}(e_x) + B_p\hat{\Lambda} R_{\Delta u_0}(\Delta u) \right]
 \geq \nonumber\\- \alpha(h(x_m))+\Delta, \eea
where $R_{\Delta u_0}(\Delta u(t))$ and $R_{e_0}(e_x(t))$ represent magnitude limited signals of $\Delta u(t)$ and $e_x(t)$, with suitable limits $\Delta u_0$ and $e_0$, respectively. The KKT condition for the QP-ZCBF safety filter is defined as:
\begin{subequations}
\begin{align}
& r - r^* - L_{B_m} h(x_m)^T \lambda  = 0 \label{KKT_r_CCRM_1}\\
& \lambda (-L_{A_m x_m} h(x_m) - L_{B_m} h(x_m) r - L_{L} h(x_m) R_{e_0}(e_x) \notag \\
& - L_{B_p\hat{\Lambda}} h(x_m) R_{\Delta u_0}(\Delta u)  -\alpha(h(x_m))+\Delta) = 0\label{KKT_r_CCRM_2}\\
& -L_{A_m x_m} h(x_m) - L_{B_m} h(x_m) r - L_{L} h(x_m) R_{e_0}(e_x) \notag\\ 
& - L_{B_p\hat{\Lambda}} h(x_m) R_{\Delta u_0}(\Delta u) - \alpha(h(x_m)) + \Delta\leq 0 \label{KKT_r_CCRM_3}
\end{align}
\end{subequations}
where $\lambda \geq 0$ is chosen so as to ensure feasibility, and $\alpha(h(x(t)))=\gamma(e_h)h(x(t))$, with $\gamma(e_h(t))$ defined as in \eqref{error_based_gamma_1}. It should be noted that \eqref{KKT_r_CCRM_2}-\eqref{KKT_r_CCRM_3} are well defined for any choice of $e_x(t)$ and $\Delta u(t)$, which are yet to be shown to be bounded.

\subsection{Adaptive control in the presence of magnitude limits}

The same adaptive controller as presented in \eqref{adaptive_feeback_controller}-\eqref{theta_r_adaption_law} is utilized here as well with a few modifications. The corresponding error dynamics between the CCRM in \eqref{CCRM} and the plant in \eqref{LinearPlantModel} can be derived to be \cite{Gibson_2013, Gaudio_2022}:
\begin{equation}
    \label{error_dyanmics_CCRM}
    \begin{split}
    \dot{e}_x(t) = (A_m - L) e_x(t) + B_p \Lambda (\Tilde{\theta}_x(t) x_p(t)\\ + \Tilde{\theta}_r(t) r_s(t)) + B_p \tilde{\Lambda}(t) \Delta u(t)
    \end{split}
\end{equation}
where $\tilde{\Lambda}(t) = \Lambda - \hat{\Lambda}(t)$ is the corresponding estimation error for $\Lambda$, and  $(A_m - L)$ is Hurwitz. 
In addition to \eqref{theta_x_adaption_law}-\eqref{theta_r_adaption_law}, we adjust the parameter $\hat{\Lambda}$ as:
\begin{equation}
    \label{lambda_r_adaption_law}
    \dot{\hat{\Lambda}}(t) = \Gamma_{\Lambda} \Delta u(t) e_x^T(t) P B_p
\end{equation}
where $\Gamma_{\Lambda}$ is positive definite. Based on \eqref{error_dyanmics_CCRM} the following Lyapunov function candidate $V$ is proposed:
\begin{equation*}
\begin{split}    
    \label{Lyapunov_function_CCRM}
    V      = &\frac{1}{2} e_x(t)^T P e_x(t) + \frac{1}{2} \operatorname{Tr}  [ \Tilde{\theta}_x(t) \Gamma_1^{-1} \Tilde{\theta}_x^T(t) \Lambda ] \\
   + &\frac{1}{2} \operatorname{Tr}  [ \Tilde{\theta}_r(t) \Gamma_2^{-1} \Tilde{\theta}_r^T(t) \Lambda ]  + \frac{1}{2} \operatorname{Tr}  [ \tilde{\Lambda}(t)  \Gamma_{\Lambda}^{-1} \tilde{\Lambda}^T(t) ]
\end{split}
\end{equation*}
From the error equation in \eqref{error_dyanmics_CCRM} and the adaptive laws in \eqref{theta_x_adaption_law}, \eqref{theta_r_adaption_law}, \eqref{lambda_r_adaption_law} and the fact that $(A_m - L)^T P +  P (A_m - L) = -Q_0$, we can show that $\dot{V} =  - \frac{1}{2}e_x^T Q_0 e_x \leq 0$ and hence $e_x(t)$ is bounded. 

Unlike the previous case, we cannot immediately conclude that $x_p(t)$ is bounded, as both $x_p(t)$ and $x_(t)m$ are affected by $\Delta u(t)$. As a result, additional arguments are needed to establish boundedness. Unlike the previous case, when control inputs are limited in magnitude, one cannot guarantee global boundedness but a domain of attraction result (see \cite{Gaudio_2022,Karason_1994,Schwager_2005}), which is briefly stated below. \\

\noindent
We introduce the following definitions:
\bena
    K_{max} &=&  \max ( \sup \| \tilde{\theta}_x \| , \sup \|  \tilde{\theta}_r \| ), \sup \|  \tilde{\Lambda} \| )\\
    \beta &=& \frac{P_B K_{max}}{\|  \theta_x^* \|  + K_{max}}
\\
    a_0 &=& \frac{\overline{u}_{min} K_{max}}{\|  \theta_x^* \|  + K_{max}}\\
x_{min} &= &\frac{3 P_B K_{max} (r_{max} +1) + 3 P_B \|  \theta_r^* \|  r_{max}}{q_{min} - 3 P_B K_{max}} + \\ &&\frac{2 P_B \overline{u}_{max}}{q_{min} - 3 P_B K_{max}}
\\
    x_{max} &=& \frac{P_B a_0}{|q_{min} - 2 P_B K_{max}|}
\\
    \overline{K}_{max} &=& \frac{q_{min} - \frac{\rho}{a_0} (3 \|  \theta_r^* \|  2 \overline{u}_{max} ) q_{min}}{3 P_B + \frac{3 \rho }{a_0} (r_{max} + 1) | q_{min} - 2 P_B \|  \theta_x^* \|   |}  \\
    &&- \frac{2 P_B \|  \theta_x^* \|  }{3 P_B + \frac{3 \rho }{a_0} (r_{max} + 1) | q_{min} - 2 P_B \|  \theta_x^* \|   |}
\eena
where 
\begin{align*}
&q_{min} = \min \operatorname{eig} (Q),\: p_{min} = \min \operatorname{eig} (P)\\
&p_{max} = \max \operatorname{eig} (P)\\
&\rho = \sqrt{\frac{p_{max}}{p_{min}}},\;\overline{u}_{max} = \max_i (u_{max,i})\\
&\overline{u}_{min} = \min_i (u_{max,i}),\;P_B = \lvert P B_p \Lambda \rvert,\;\lambda_{min} = \min (\operatorname{eig}(\Lambda))\\
&\gamma_{max} = \max (\operatorname{eig}(\Gamma_x), \operatorname{eig}(\Gamma_r), \operatorname{eig} (\Gamma_\lambda))
\end{align*}
with all defined norms are Euclidean norms and the matrix $P_B$ is the induced matrix norm, which has the property $\lvert P B_p \Lambda x \rvert \leq P_B \lvert x \rvert $. Based on the introduced variables we can state the following theorem.
\begin{theorem}\cite{Schwager_2005}
\label{stability_mlac}
The plant described in \eqref{LinearPlantModel}, with the adaptive feedback controller \eqref{adaptive_feeback_controller} using the adaptive laws defined in \eqref{theta_x_adaption_law} - \eqref{lambda_r_adaption_law} has bounded trajectories for $\forall t \geq t_0$ if
\begin{itemize}
    \item [1)] $\lvert x_p (t_0) \rvert < \frac{x_{max}}{\rho}$
    \item[2)]$\sqrt{V(t_0)} < \overline{K}_{max} \sqrt{\frac{\lambda_{min}}{\gamma_{max}}}$
\end{itemize}
\end{theorem}
Since $\lvert x_p(t) \rvert < x_{max}$ for $\forall t \geq t_0$, we can state that the error variable is of the same order as the difference between saturated input $R(u(t))$ and the unsaturated $u(t)$, stated as:
\begin{equation*}
    \|  e_x(t) \| = \mathcal{O}[ \sup_{\tau \leq t} \| \Delta u(\tau) \|]
\end{equation*}
We refer to \cite{Gaudio_2022,Karason_1994,Schwager_2005} for details of the proof.

\subsection{Safety in the presence of uncertainties and control input limits}
\label{Safety in the presence of uncertainties} 
With stability guaranteed from the discussions above, we
now derive conditions for the safety of the proposed adaptive controller with magnitude saturation. We again set $\alpha(h(x))=\gamma h(x)$, where $\gamma$ is a positive constant. The goal once again is to ensure safety, that is, for inequality \eqref{eq:safety} to be satisfied. Unlike \eqref{CBF_reference_model}, we note that the modified QP-ZCBF filter in \eqref{CCRM} implies that the following inequality holds:
\begin{equation}
    \label{CBF_reference_model2}
    \begin{split}
        \dot{h}_m   &= \frac{\partial h}{\partial x}|_{x_m} 
        [ A_m x_m(t) + B_m r_s(t) + L R_{e_0}(e_x(t))\\
        &+ B_p \hat{\Lambda} R_{\Delta u_0}(\Delta u(t))
        ]\geq -\gamma(e_h(t)) h(x_m(t))+\Delta
    \end{split}
\end{equation}
using $\Delta e_x(t) = R_{e_0}(e_x(t)) - e_x(t)$ and $\bar{\Delta} u(t) = R_{\Delta u_0}(\Delta u(t)) - \Delta u(t)$ we can reformulate
\begin{equation}
    \label{CBF_reference_model2b}
    \begin{split}
        \dot{h}_m   &= \underbrace{\frac{\partial h}{\partial x}|_{x_m}}_{c_0} 
        [ \underbrace{A_m x_m(t) + B_m r_s(t)}_{c_1} + \underbrace{L e_x(t)}_{c_2}  \\
        &+ \underbrace{B_p \hat{\Lambda}(t) \Delta u(t)}_{c_3} + \underbrace{L \Delta e_x(t) + B_p \hat{\Lambda}(t) \bar{\Delta} u}_{S_{\Delta}}
        ]\\
        &\geq -\underbrace{\gamma(e_h(t))  h(x_m(t))}_{c_4}+\Delta
    \end{split}
\end{equation}
We can derive $\dot h_p$ using \eqref{LinearPlantModel} and considering \eqref{magnitude_limits} as
\begin{equation}
    \label{CBF_plant_adaptive2}
    \begin{split}
    \dot{h}_p = \frac{\partial h}{\partial x}|_{x_p}
    [A_p x_p &+ B_p \Lambda ( \hat{\theta}_x(t) x_p(t) \\
    &+ \hat{\theta}_r(t) r_s(t) + \Delta u(t)) ]
    \end{split}
\end{equation}
using \eqref{LinearPlantModel} and the error between $x_p$ and $x_m$, we can state
\begin{equation}
\label{hp_do_c1_es}
\begin{split}
\dot{h}_p = 
\underbrace{\frac{\partial h}{\partial x}|_{x_p}}_{d_0}
[
&\underbrace{A_m x_m(t) + B_m r_s(t)}_{c_1}
+ \underbrace{L e_x(t)}_{c_2}\\
&+ \underbrace{B_p \hat{\Lambda}(t) \Delta u(t)}_{c_3} + \Bar{e}_\Delta
]
\end{split}
\end{equation}
with $\bar{e}_{\Delta} = \bar{e} - L e_x(t) + B_p \Tilde{\Lambda}(t) \Delta u(t)$. Algebraic manipulations allow us to rewrite \eqref{hp_do_c1_es} as 
\begin{equation}
    \label{CBF_deriation_inequality2s}
    \begin{split}
    \dot{h}_p &= d_0 [c_1 + c_2 + c_3 + \Bar{e}_{\Delta}]\\ 
    &\geq - c_4 +\Delta+  \underbrace{c_0 \Bar{e}_{\Delta} + (d_0 - c_0) (c_1 + c_2 + c_3 + \Bar{e}_{\Delta})}_{w(t)}-c_0 S_{\Delta}
    \end{split}
\end{equation} 
Noting that the inequality in \eqref{CBF_deriation_inequality2s} is very similar to \eqref{CBF_deriation_inequality}, similar relations to \eqref{lipschitz_gradient} and \eqref{lipschitz_barrierfunction} can be employed to derive an inequality
\begin{equation}
    \label{lower_bound_h_dot2}
    \begin{split}
    \dot{h}_p \geq - \gamma(e_h(t))  h(x_p(t)) +\Delta- \bar{F}_{\Delta}(|\Bar{e}_{\Delta}|, | e_x |)
    \end{split}
\end{equation}
where
\begin{equation}
\begin{split}
   \bar{F}_{\Delta}(|\Bar{e}_{\Delta}|, &| e_x(t) |) =  \gamma\kappa_4 | e_x | + |c_0| |\Bar{e}_{\Delta}|\\
   & +\kappa_3 |e_x | \begin{pmatrix}  |c_1| + |c_2| +|c_3| + |\Bar{e}_{\Delta}| \end{pmatrix} -  C_0 S_{\Delta}
\end{split}
\end{equation}
Using similar arguments that utilize the asymptotic convergence of $ \bar{F}(t)$ to zero, we obtain once again that the closed-loop adaptive system will remain safe for all $t\geq t_0$ if
\begin{equation}
h(x_p(t))\geq h_0 \,\,\,\,\, \forall t\geq [t_0,t_0+T]
\label{eq:safetycondn2}
\end{equation}
where $\gamma h_0 \geq F_{{\rm max}}$,where 
\begin{equation*}
F_{{\rm max}} = \max_{t\in [t_0,t_0+T]}\bar{F}(t)
\end{equation*}
An EBR as in (36) can be introduced as in Section III to allow condition (52) to be satisfied for a large class of $h(x(t))$. 

A theorem similar to Theorem \ref{Theorem_Safe_Basic_Adaptive_Controller} can be derived to encapsulate the safety property of the closed-loop adaptive system with magnitude saturation using the controller specified by \eqref{adaptive_feeback_controller}-\eqref{theta_r_adaption_law} and \eqref{lambda_r_adaption_law}. The resulting closed-loop system therefore is stable, safe, and accommodates magnitude constraints on the control input. 

\section{Simulations}
\label{Simulation}
Two simulation examples are provided in this section to illustrate the properties of the safe and stable adaptive control approach described in this paper that includes a QP-CBF filter and an EBR based damping term $\gamma$.
\subsection{Obstacle avoidance}
The proposed controller is applied to a simple 2D obstacle avoidance problem in the simulation case here. The used model is defined as:
\begin{equation*}
\begin{bmatrix}
\dot{x}\\
\dot{y}
\end{bmatrix} =
    \begin{bmatrix}
       -2 &   0\\
       0 &  -2\\
    \end{bmatrix}
\begin{bmatrix}
x\\
y
\end{bmatrix}
+
\begin{bmatrix}
   0.8 &   0\\
   0 &  0.8\\
\end{bmatrix}
\begin{bmatrix}
u_1\\
u_2
\end{bmatrix}
\end{equation*}
where $x$ and $y$ represent the position states, $\dot{x}$ and $\dot{y}$ are their derivatives. The control inputs being represented by $u_1$ and $u_2$. The magnitude limit on both control inputs $u_0$ was set to $10$. The reference model is defined as:\vspace{0.2cm}
\begin{minipage}{.45\linewidth}
\begin{equation*}
    A_m = 
    \begin{bmatrix}
       -1 &   0\\
       0 &  -1\\
    \end{bmatrix},
\end{equation*}
\end{minipage}%
\begin{minipage}{.45\linewidth}
\begin{equation*}
    B_m = 
    \begin{bmatrix}
       -1 &   0\\
       0 &  -1\\
    \end{bmatrix}
\end{equation*}
\end{minipage}\vspace{0.2cm}

\noindent To ensure safety in the context of obstacle avoidance, we choose for each obstacle the following CBF constraint:
\begin{equation*}
    h_o = (x - x_o)^2 + (y - y_o)^2 - r_o^2 \geq 0
\end{equation*}
where $x_o$ and $y_0$ represent the $x$ and $y$ position of the obstacles center and $r_o$ the radius of the circular obstacles.
\begin{figure}[h!]
\centering
\includegraphics[width=1\columnwidth]{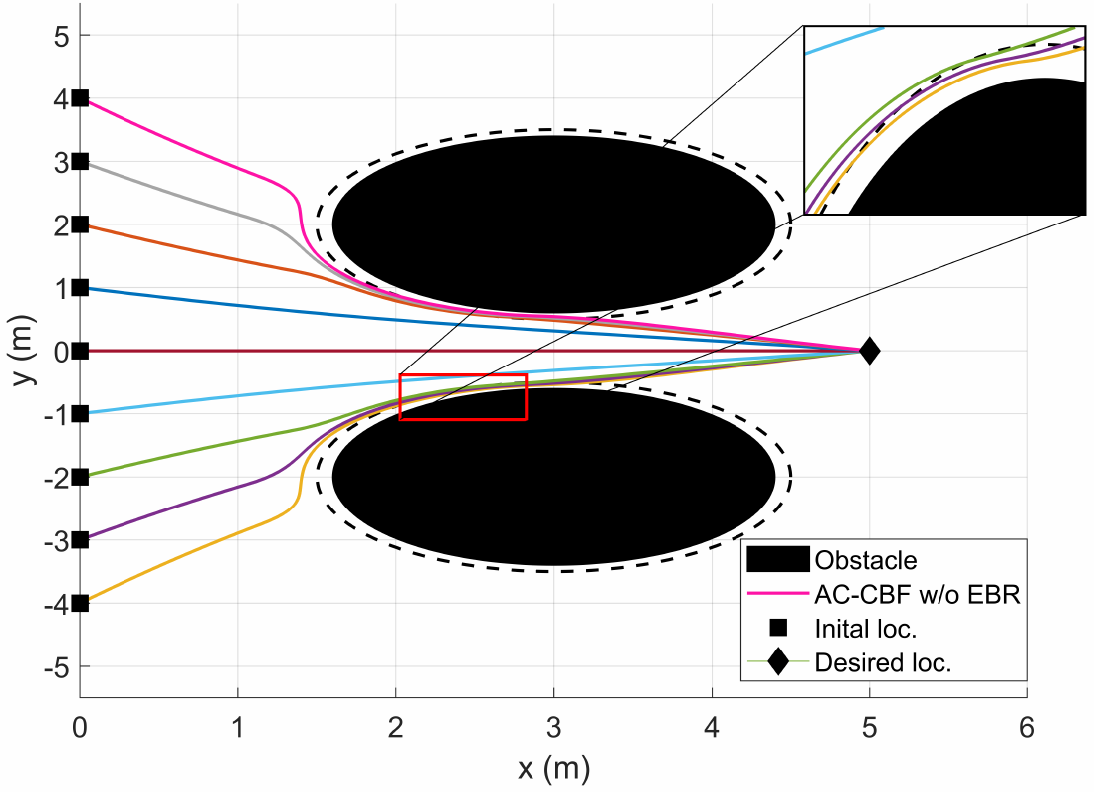}
\caption{Comparison of state trajectories with different initial states using an AC-CBF without EBR.}
\label{fig:ObstacleAvoidanceNoERB}
\end{figure}
\begin{figure}[h!]
\centering
\includegraphics[width=0.85\columnwidth]{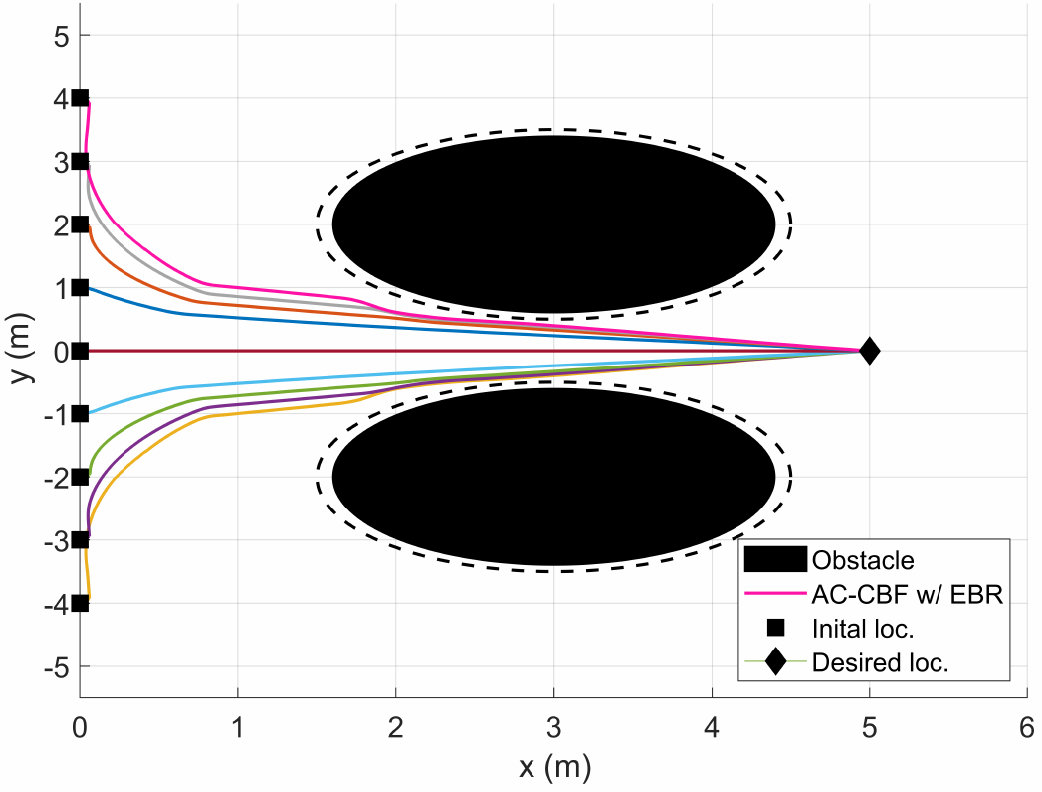}
\caption{Comparison of state trajectories with different initial states using a AC-CBF with EBR.}
\label{fig:ObstacleAvoidanceERB}
\end{figure}

Figure \ref{fig:ObstacleAvoidanceNoERB} shows the state trajectories of the discussed adaptive control framework without the proposed error-based relaxation (EBR) of the CBF constraint for adaptive closed-loop systems with different initial locations. It is apparent that even though the adaptive controller learns the parameter, it cannot ensure safety since it cannot learn fast enough. This leads to a safety violation for most of the shown trajectories. Figure \ref{fig:ObstacleAvoidanceERB} shows the state trajectories for systems with different initial locations using the proposed adaptive controller with EBR of the CBF constraint. It can be seen that the proposed method introduces a helpful conservatism where the model is not yet learned. This renders the system safer and ensures that no trajectory violates the safety constraint even though the model is not learned yet.
\subsection{Missile pitch dynamics}
The used missile pitch dynamics model is shared in \cite{Wise1991}. The model was modified such that a model mismatch can be considered:
\begin{equation*}
\begin{split}
\begin{bmatrix}
\dot{\alpha}\\
\dot{q}
\end{bmatrix} =
&\Delta A_p\begin{bmatrix}
Z_{\alpha} & Z_{q}\\
M_{\alpha} & M_{q}
\end{bmatrix}
\begin{bmatrix}
\alpha\\
q
\end{bmatrix}
+
\begin{bmatrix}
Z_{\delta}\\
M_{\delta}
\end{bmatrix}
\lambda_{\delta}\delta \\
= &\Delta A_p \begin{bmatrix}
-0.8757 & 1\\
-68.9210 & 0
\end{bmatrix}
\begin{bmatrix}
\alpha\\
q
\end{bmatrix}
+
\begin{bmatrix}
-0.1531\\
-74.2313
\end{bmatrix}
\lambda_{\delta}\delta
\end{split}
\end{equation*}
where $\alpha$ represents the aerodynamic angle of attack (AoA), $q$ the pitch rate and $\delta$ the fin deflection. $\Delta A_p$ and $\lambda_{\delta}$ are scalar parameters used to introduce static parameter deviations. For the here regarded case, the parameters were set to $\Delta A_p=1.2$ and $\lambda_{\delta}=0.6$. The regarded linear dynamics are modeled at Mach 0.8 and an altitude of 4000 ft, with a trim angle of attack of 6 degrees. The magnitude limits on the control input are set to 10 degrees for $\delta$. The reference model of the system was chosen by defining a nominal closed-loop response, using a conventional LQR technique to define a suitable feedback controller \cite{Kalman_1960LQR}. The reference model is defined as:

\begin{minipage}{0.45\linewidth}
\begin{equation*}
    \small 
    A_m = \begin{bmatrix}
       -0.8707 &   0.9927\\
      -65.5877 &  -3.5903\\
    \end{bmatrix},
\end{equation*}\
\end{minipage}
\hfill
\begin{minipage}{0.45\linewidth}
\begin{equation*}
    \small 
    B_m = \begin{bmatrix}
    0.1395 &  -0.0364\\
    68.2893 & -17.7947\\ 
    \end{bmatrix}
\end{equation*}
\end{minipage}\vspace{0.2cm}
For the here presented simulation case, it was chosen that the maximum missile's AoA is limited by an arbitrary value. To ensure that, we choose the following CBF constraint:
\begin{equation*}
    h_{\alpha} = \alpha_{max} - \alpha \geq 0
\end{equation*}
\begin{figure}
\centering
\includegraphics[width=0.85\columnwidth]{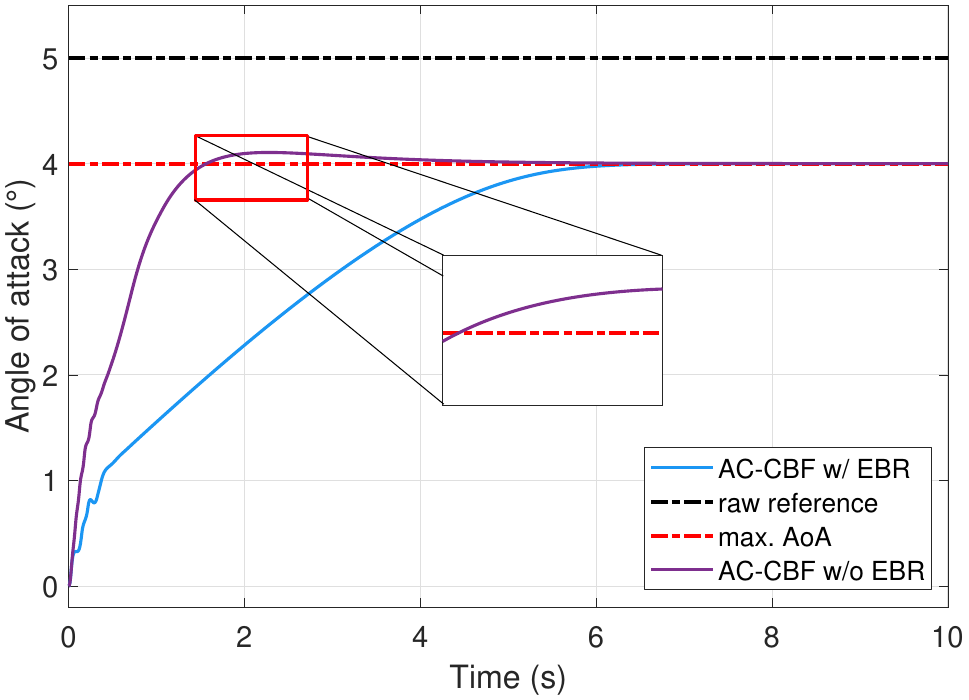}
\caption{Trajectory comparison of state AoA for AC-CBF with EBR and without EBR w.r.t. to the commanded and the maxmimum AoA.}
\label{fig:AoAComparison}
\end{figure}
\begin{figure}
\centering
\includegraphics[width=0.85\columnwidth]{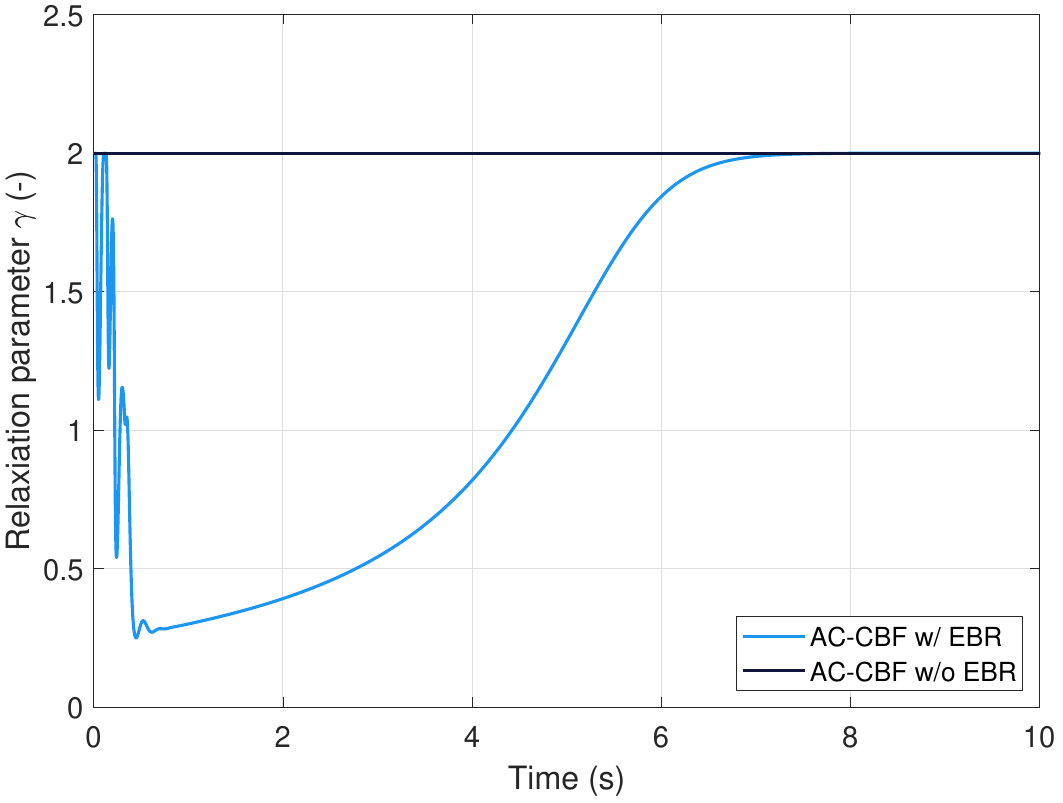}
\caption{Time history of relaxation parameter $\gamma$ for the case with EBR and without EBR.}
\label{fig:GammaComparison}
\end{figure}
\begin{figure}
\centering
\includegraphics[width=0.85\columnwidth]{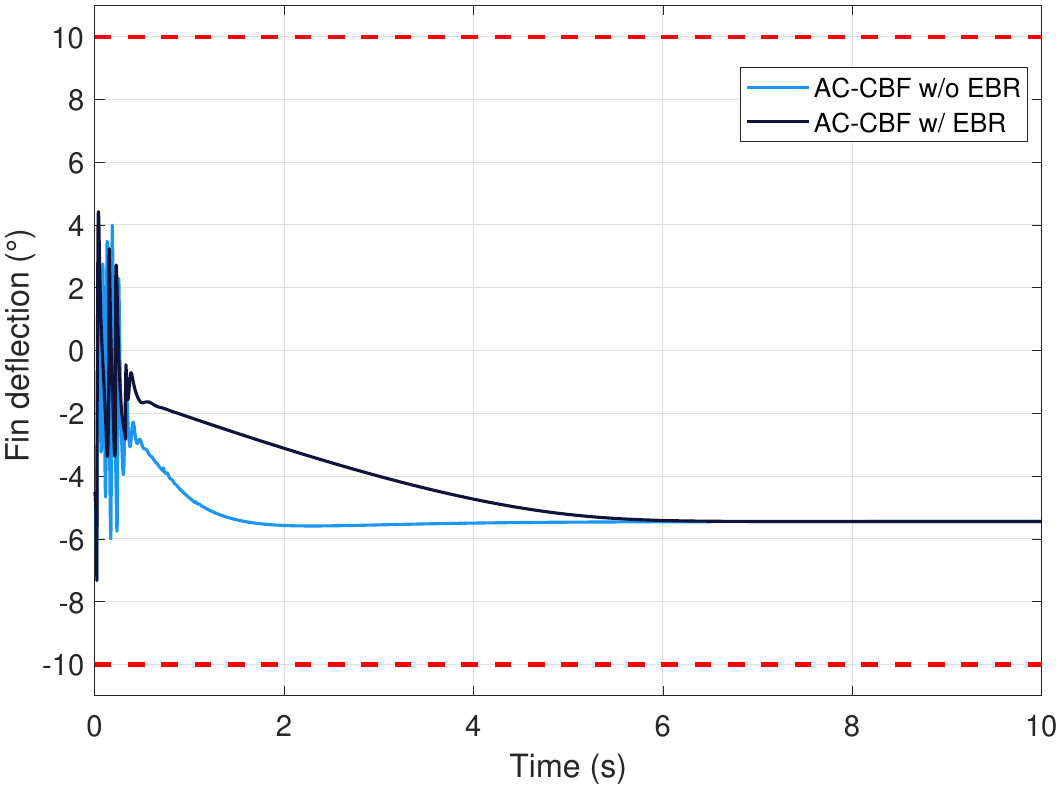}
\caption{Comparison of the time history of the fin deflection $\delta$ for the adaptive controller with EBR and without EBR.}
\label{fig:ControlInputComparison}
\end{figure}
Figure \ref{fig:AoAComparison} compares the closed-loop response of the suggested adaptive controller in both versions, with and without the error-based relaxation parameter. In this case, the desired AoA is set to $5$ degrees, but the maximum allowable state was set to $4$ degrees. It can be seen that the controller without the EBR violates the defined maximum AoA constraint. The controller with EBR is able to cope better with the problem of uncertainties and uses the added conservatism during the learning to converge to the defined maximum value when the confidence in the model allows it. Figure \ref{fig:GammaComparison} shows how the relaxation parameter $\gamma (e)$, for the adaptive controller with EBR, decreases at the beginning when the model is the least known. During the learning, the value increases and indicates a higher confidence in the operation within the safe set. Figure \ref{fig:ControlInputComparison} compares the time history of the control input for the proposed adaptive controller with and without EBR, which shows that the control input always stays well within its limits. We expect that the oscillations can be improved further by deploying rate limits and extending the proposed approach in this paper along the lines of \cite{Gaudio_2022}.

\section{Conclusions}
\label{Conclusions}
This paper takes a step in combining adaptive control methods with safety-critical methods for a specific class of dynamic systems. In addition to ensuring safety and stability, the proposed adaptive control design also seeks to accommodate magnitude constraints on the control input. The proposed approach employs a combination of classical adaptive control, closed-loop reference model that is calibrated to accommodate input saturation, a quadratic programming based CBF filter, and an error-based relaxation of the damping characteristics of the CBF. The resulting combination is shown to lead to global boundedness without input constraints, and a domain of attraction result with input constraints. In both cases, conditions for the existence of a control barrier function are derived. Numerical results validate the analytical derivations.

\bibliographystyle{IEEEtran}
\bibliography{references} 

\appendices

  \section{Extended derivation of safety conditions for adaptive controller in the presence of uncertainties}
  \label{FirstAppendix}
  We assume the following CBF for the plant:
\begin{equation}
\label{eq:safety_extended}
\dot h_p\geq -\gamma h(x_p)
\end{equation}
The integrated QP-CBF filter ensures that the following  inequality holds
\begin{equation}
    \label{CBF_reference_model_extended}
    \dot{h}_m   = \underbrace{\frac{\partial h}{\partial x}|_{x_m}}_{a_0} 
    [ \underbrace{A_m x_m + B_m r_s}_{a_1}
    ]\geq -\underbrace{\gamma h(x_m)}_{a_2} +\Delta
\end{equation}
In order to ensure that a ZCBF exists for the adaptive system specified by \eqref{LinearPlantModel},\eqref{adaptive_feeback_controller}-\eqref{theta_r_adaption_law}, we consider
\begin{equation}
    \label{CBF_plant_adaptive}
    \begin{split}
    \dot{h}_p = \frac{\partial h}{\partial x}|_{x_p}
    \begin{bmatrix}
    A_p x_p + B_p \Lambda ( \widehat{\theta}_x x_p + \widehat{\theta}_r r_s ) \end{bmatrix}
    \end{split}
\end{equation}
From \eqref{LinearPlantModel}, \eqref{matching_condition_1}-\eqref{adaptive_feeback_controller}, \eqref{eq:idealu},  \eqref{CBF_reference_model_extended} and the control input error $e_u=u-u^* = \widetilde{\theta}_x x_p + \widetilde{\theta}_r r_s$, we obtain that
\begin{equation}
    \label{hp_bo_a1_e}
    \begin{split}
    \dot{h}_p = 
    \frac{\partial h}{\partial x}|_{x_p}
    [ A_p x_p + B \Lambda (\widetilde{\theta}_x x_p + \widetilde{\theta}_r r_s +   \theta^*_x x_p +   \theta^*_r r_s)
    ]
    \end{split}
\end{equation}
with additionally applying \eqref{matching_condition_1} and \eqref{matching_condition_2} from assumption \ref{assumption_matching_condition}, the definition of the tracking error $e_x=x_p-x_m$, we obtain
\begin{equation}
    \label{hp_bo_a1_e_extended}
    \begin{split}
    \dot{h}_p = 
    \underbrace{\frac{\partial h}{\partial x}|_{x_p}}_{b_0}
    [
    \underbrace{A_m x_m + B_m r_s}_{a_1}
     + \underbrace{A_m e_x + B_p \Lambda e_u}_{\Bar{e}}
    ]
    \end{split}
\end{equation}
Algebraic manipulations allow us to rewrite \eqref{hp_bo_a1_e_extended} using \eqref{CBF_reference_model_extended} as
\begin{equation}
    \label{CBF_deriation_inequality_extended}
    \begin{split}
    \dot{h}_p &= b_0 [a_1 + \Bar{e}]\\ 
    &\geq - a_2 +\Delta+  \underbrace{a_0 \Bar{e} + (b_0 - a_0)\Bar{e} + (b_0 - a_0) a_1}_{z(t)}
    \end{split}
\end{equation}
Since the goal is to establish safety of the closed-loop adaptive system, we utilize the following two inequalities:
\begin{equation}
\label{lipschitz_gradient_extended}
    \begin{split}
    | g(x_p) - g(x_m) | \leq \kappa_1 | e_x |
    \end{split}
\end{equation}
\begin{equation}
\label{lipschitz_barrierfunction_extended}
    \begin{split}
    | h(x_p) - h(x_m) | \leq \kappa_2 | e_x |
    \end{split}
\end{equation}
where $g(x)=\frac{\partial h}{\partial x}$, $\kappa_1$ and $\kappa_2$ are Lipschitz constants associated with $g(x_p)$ and $h(x_p)$, respectively. We note additionally from Theorem \ref{Theorem_Basic_Adaptive_Controller} that $\|x_p(t)\|$, $\|h(x_p)\|$ and $\|\Bar{e}(t)\|$ are bounded. Therefore, $|z(t)|\leq z_0$, where $z_0$ is defined as
\begin{equation}
    z_0= |a_0| |\Bar e| + \kappa_1 |e_x | \begin{pmatrix} |\Bar e| + |a_1| \end{pmatrix}
\label{eq:bound1}
\end{equation} 
with $- z_0 \leq  z(t) \leq z_0$ holding. Using the lower bound $-z_0$ for $z(t)$, we rewrite \eqref{CBF_deriation_inequality_extended} as
\begin{equation}
    \label{lower_bound_h_dot_extended}
    \begin{split}
    \dot{h}_p \geq - \gamma h(x_p) +\Delta- \bar{F}(|\Bar{e}|, | e_x |)
    \end{split}
\end{equation}
where
\begin{equation}
   \bar{F}(|\Bar{e}|, | e_x |) =  \gamma\kappa_2 | e_x | + |a_0| |\Bar e| + \kappa_1 |e_c | \begin{pmatrix} |\Bar e| + |a_1|\end{pmatrix}
\end{equation}
The inequality in \eqref{lower_bound_h_dot} implies that safety of the closed-loop adaptive system will be guaranteed after $t\geq t_0+T$, where $T$ is a finite interval, as $\Bar{F}(t)\rightarrow 0$ as $t\tends \infty$, and therefore $|\Bar{F}(t)|\leq \Delta\; \forall t \geq t_0+T$. This in turn implies that the closed-loop adaptive system will remain safe for all $t\geq t_0$ if
\begin{equation}
h(x_p(t))\geq h_0 \,\,\,\,\, \forall t\geq [t_0,t_0+T]
\label{eq:safetycondn}
\end{equation}
where $\gamma h_0 \geq F_{{\rm max}}$,where 
\begin{equation}
\label{eq:safetycondn_F}
F_{{\rm max}} = \max_{t\in [t_0,t_0+T]}\bar{F}(t)
\end{equation}
As $F(t)$ is bounded, it is clear that such an $F_{{\rm max}}$ exists. Condition \eqref{eq:safetycondn} is satisfied if there is a separation between the period of adaptation and the time at which the system approaches its limit of safety. This property is summarized in the following theorem, where
  
  \section{Extended derivation of safety conditions for adaptive controller in the presence of uncertainties and control input limits}
  \label{SecondAppendix}
  Basd on the CCRM defined in \eqref{CCRM} the following inequality holds:
\begin{equation}
    \label{CBF_reference_model2_extended}
    \begin{split}
        \dot{h}_m   = \frac{\partial h}{\partial x}|_{x_m} 
        [ &A_m x_m + B_m r_s + L R_{e_0}(e_x)\\
        &+ B_p \widehat{\Lambda} R_{\Delta u_0}(\Delta u)
        ]\geq -\gamma(e_h) h(x_m)+\Delta
    \end{split}
\end{equation}
using $\Delta e_x = R_{e_0}(e_x) - e_x$ and $\bar{\Delta} u = R_{\Delta u_0}(\Delta u) - \Delta u$ we can reformulate
\begin{equation}
    \label{CBF_reference_model2b}
    \begin{split}
        \dot{h}_m   = \underbrace{\frac{\partial h}{\partial x}|_{x_m}}_{c_0} 
        [ &\underbrace{A_m x_m + B_m r_s}_{c_1} + \underbrace{L e_x}_{c_2}  + \underbrace{B_p \widehat{\Lambda} \Delta u}_{c_3} \\
        &+ \underbrace{L \Delta e_x + B_p \widehat{\Lambda} \bar{\Delta} u}_{S_{\Delta}}
        ]\geq -\underbrace{\gamma(e_h)  h(x_m)}_{c_4}+\Delta
    \end{split}
\end{equation}
We can derive $\dot h_p$ using \eqref{LinearPlantModel} and considering \eqref{magnitude_limits} as
\begin{equation}
    \label{CBF_plant_adaptive2}
    \begin{split}
    \dot{h}_p = \frac{\partial h}{\partial x}|_{x_p}
    \begin{bmatrix}
    A_p x_p + B_p \Lambda ( \widehat{\theta}_x x_p + \widehat{\theta}_r r_s + \Delta u) \end{bmatrix}
    \end{split}
\end{equation}
with applying \eqref{matching_condition_1} and \eqref{matching_condition_2} from assumption \ref{assumption_matching_condition}, the control input error $e_u=u-u^* = \widetilde{\theta}_x x_p + \widetilde{\theta}_r r_s$ and the error $e_x=x_p-x_m$, we can state
\begin{equation}
\label{hp_do_c1_e}
\begin{split}
\dot{h}_p = 
\underbrace{\frac{\partial h}{\partial x}|_{x_p}}_{d_0}
[
&\underbrace{A_m x_m + B_m r_s}_{c_1}
+ \underbrace{L e_x}_{c_2}\\
&+ \underbrace{B_p \widehat{\Lambda} \Delta u}_{c_3} + \Bar{e}_\Delta
]
\end{split}
\end{equation}
with $\bar{e}_{\Delta} = \bar{e} - L e_x + B_p \widetilde{\Lambda} \Delta u$ and $\widetilde{\Lambda} = \Lambda - \widehat{\Lambda}$. Algebraic manipulations allow us to rewrite \eqref{hp_do_c1_e} as 
\begin{equation}
    \label{CBF_deriation_inequality2}
    \begin{split}
    \dot{h}_p &= d_0 [c_1 + c_2 + c_3 + \Bar{e}_{\Delta}]\\ 
    &\geq - c_4 +\Delta+  \underbrace{c_0 \Bar{e}_{\Delta} + (d_0 - c_0) (c_1 + c_2 + c_3 + \Bar{e}_{\Delta})}_{w(t)}-c_0 S_{\Delta}
    \end{split}
\end{equation}
Noting that the inequality in \eqref{CBF_deriation_inequality2} is very similar to \eqref{CBF_deriation_inequality_extended}, similar relations to \eqref{lipschitz_gradient_extended} and \eqref{lipschitz_barrierfunction_extended} can be employed to derive an inequality
\begin{equation}
    \label{lower_bound_h_dot2}
    \begin{split}
    \dot{h}_p \geq - \gamma(e_h)  h(x_p) +\Delta- \bar{F}_{\Delta}(|\Bar{e}_{\Delta}|, | e_x |)
    \end{split}
\end{equation}
where
\begin{equation}
\begin{split}
   \bar{F}_{\Delta}(|\Bar{e}_{\Delta}|, &| e_x |) =  \gamma\kappa_4 | e_x | + |c_0| |\Bar{e}_{\Delta}|\\
   & +\kappa_3 |e_x | \begin{pmatrix}  |c_1| + |c_2| +|c_3| + |\Bar{e}_{\Delta}| \end{pmatrix} -  C_0 S_{\Delta}
\end{split}
\end{equation}
Using similar arguments that utilize the asymptotic convergence of $ \bar{F}(t)$ to zero, we obtain once again that the closed-loop adaptive system will remain safe for all $t\geq t_0$ if
\begin{equation}
h(x_p(t))\geq h_0 \,\,\,\,\, \forall t\geq [t_0,t_0+T]
\label{eq:safetycondn2}
\end{equation}
where $\gamma h_0 \geq F_{{\rm max}}$,where 
\begin{equation*}
F_{{\rm max}} = \max_{t\in [t_0,t_0+T]}\bar{F}(t)
\end{equation*}
An EBR as in (36) can be introduced as in Section III to allow condition (52) to be satisfied for a large class of $h(x)$. 

\end{document}